\documentclass{jpsj2}

\newcommand{\nc}{\newcommand}
\nc{\rnc}{\renewcommand}

\nc{\nn}{\nonumber}
\nc{\ds}{\displaystyle}

\nc{\ch}{\cosh}
\nc{\varph}{\sinh}
\nc{\sh}{\sinh}
\rnc{\th}{\tanh}

\nc{\lam}{\lambda}
\nc{\gam}{\gamma}
\rnc{\d}{\Delta}

\nc{\feta}{\frac{\eta}{2}}
\nc{\fgam}{\frac{\gam}{2}}
\nc{\ifgam}{\frac{i\gam}{2}}
\nc{\p}{\Psi\Psi^{\ast}}

\nc{\inv}{^{-1}}
\nc{\sm}{$-$}

\nc{\bra}{\langle}
\nc{\ket}{\rangle}

\nc{\vac}{\mid0\,\ket}
\nc{\vacc}{\bra\,0\mid}

\nc{\sumir}{\sum^r_{i=1}}
\nc{\sumiN}{\sum^N_{i=1}}
\nc{\sumnN}{\sum^N_{n=1}}
\nc{\summr}{\sum^r_{m=1}}
\nc{\prodiN}{\prod^N_{i=1}}
\nc{\prodjN}{\prod^N_{j=1}}
\nc{\prodkN}{\prod^N_{k=1}}

\rnc{\(}{\left(}
\rnc{\)}{\right)}
\rnc{\[}{\left[}
\rnc{\]}{\right]}

\nc{\xxx}{$XXX \,$}

\title
{
Evaluation of Dynamic spin structure factor\\
for the spin-1/2 XXZ chain in a magnetic field
}

\author
{ 
Jun Sato, Masahiro Shiroishi and Minoru Takahashi
}

\inst
{
Institute for Solid State Physics, University of Tokyo, Kashiwa, 
Chiba 277-8581, Japan
}

\recdate{\today}

\abst{
Transition rates and dynamic spin structure factor at zero temperature 
for the spin-1/2 XXZ chain at critical regime in a magnetic field
are numerically evaluated in terms of the exact determinant representations for 
the form factors and norms of the Bethe eigenstates.
We have seen that the transition rates converges toward the constant
function with the value $1$ in the limit $\d\rightarrow0$.
The observed critical exponent of the singularity at the lower boundary
is compared with the one predicted
from the comformal field theory.
We confirm that they are in good agreement.
Further we have discovered that a small peak emerges near the upper boundary
in the line shape of $S(q,\omega)$ for $0<\d<1$.
}

\kword{
Bethe ansatz, XXZ model, dynamic spin structure factor, CFT
}

\begin{document}

\maketitle

\section{Introduction}

Since Bethe\cite{Bethe} in 1931 formulated the method (Bethe ansatz)
 for obtaining the exact 
eigenvalues and eigenvectors of the one-dimensional spin-1/2 Heisenberg model,
it has been extensively
generalized
to study the excitations and the thermodynamics of
the model exactly.\cite{Takabook}
However, from the point of view of the dynamics of Heisenberg antiferromagnetic chains,
only approximate results had been obtained until recently,
partly due to the lack of the exact expression for the form factors.

Heisenberg antiferromagnetic chains are realized in quasi-one-dimensional magnetic
insulators, such as,
$\mathrm{Cu(C_4H_4N_2)(NO_3)_2}$ \cite{cuno}, and $\mathrm{KCuF_3}$ \cite{kcu}.
Dynamic spin structure factor $S(q,\omega)$, which is the Fourier transform of 
dynamical correlation functions, is of considerable importance,
since they
are directly comparable with inelastic neutron scattering experiments
of these quasi one-dimensional substances.\cite{stone}

In 1967 Niemeijer obtained an exact expression of $S(q,\omega)$
for the spin-1/2 XY model at any temperature\cite{nmj}.
It is the special case ($\d=0$) of XXZ model,
where all the correlation functions can be calculated by Jordan-Wigner
transformation.
Unfortunately
this exact calculation can not be extended for general case of $\d\neq 0$,
and only approximate evaluations of $S(q,\omega)$ had been attempted.\cite{beck}
It was worth noting, therefore, that an analytical result for the 
two-spinon dynamic spin structure factor for the spin-1/2 massive XXZ model ($\d\geq 1$) 
in zero magnetic field was obtained \cite{bg,bgkar1,bgkar2}
from the multiple integral representation for form factors\cite{jimbo},
which is based on the infinite dimensional symmetries of the quantum affine algebra 
$U_q(\hat{sl}(2,{\bf C}))$.

However, this analytical calculation is not applicable
to the case with nonzero magnetic field.
Karbach et al. classified the dynamically dominant excitations of the spin-1/2 XXX
Heisenberg model ($\d=1$)
in a magnetic field.
They subsequently calculated their transition rates
pertaining to dynamic spin structure factor,
directly from 
the Bethe wave functions for finite chains with up to $N=32$ sites\cite{line,quasi}.
Although
solving the Bethe ansatz equations in its own is possible for
$N\approx 10^3$,
constructing the Bethe wave functions is limited to
$N\approx 32$.
The problem of this direct calculation lies in 
the evaluation of the sum over the $r!$ magnon permutations in the coefficients of the
coordinate Bethe wave functions, where $r$ is the number of the down spins.

Quite recently,
this problem was solved by the novel work of Kitanine et al.,
who derived
the exact determinant representation
for the form factors of local spin operators for arbitrary finite systems.
\cite{kitanine}
In the framework of
the algebraic Bethe ansatz\cite{Korepinbook},
they have found that
any local operators can be written as the elements of
the quantum monodromy matrix.
%
%
Consequently
the calculation of form factors reduces to
that of scalar products between a Bethe state and an arbitrary state,
which has a determinant representation.
Then
it enables us to calculate transition rates directly from the solution
of Bethe ansatz equations,
without constructing the Bethe wave functions.

Biegel et al. calculated dynamic spin structure factors
for chains with the order $N\approx 10^3$
using this determinant
representation in the case of XXX Heisenberg chain in a magnetic field\cite{karxxx}
and in the case of XXZ Heisenberg chain 
at critical regime in zero magnetic field\cite{karxxz,karxx0}.
As much as these calculations are still for the finite systems,
we can deal with much larger system sizes than in the past.

In this paper,
as an extension of these works,
we shall generalize the work by Biegel et al. and
evaluate dynamic spin structure factors
for the spin-1/2 XXZ chain in a magnetic field. Under the existance of a
magnetic field, only the XXX case has been dealt with so far.

\section{Bethe ansatz}

Let us consider the one-dimensional spin-1/2 XXZ model
with periodic boundary conditions
in a magnetic field
\begin{equation}
\label{ham}
H=J\sum^N_{n=1}\left\{S_n^xS^x_{n+1}+S_n^yS^y_{n+1}
    +\Delta (S_n^zS^z_{n+1}-\dfrac{1}{4})\right\}
    -h\sumnN S_n^z,
\end{equation}
where $J>0$ is the coupling constant,
$S^{x,y,z}=\dfrac{1}{2}\sigma^{x,y,z}$,
$\d$ is the anisotropy parameter
and $h$ is the external magnetic field applied to the positive direction of $z$-axis.
$\sigma^{x,y,z}$ represents the standard Pauli matrices.

It is exactly solved by Bethe ansatz method.
Bethe eigenstates with $r$ down spins are constructed by a set of rapidities
$\{z_1,\dots,z_r\}$,
which is a solution of Bethe ansatz equations
\begin{eqnarray}
\label{bae}
N\tan\inv[\cot\frac{\gam}{2}\tanh z_i]
&=\pi I_i+\sum^r_{j\neq i}\tan\inv[\cot\gam\th(z_i-z_j)],\nn\\
&i=1,\dots,r,
\end{eqnarray}
where the parameter $\gam$ is related to the anisotropy $\d$ as
$\gam=\cos\inv\d$.
%
%
$I_i$ are Bethe quantum numbers,
which have integer values for odd $r$ and
half odd integer values for even $r$.
The total momentum and energy eigenvalues are given by
\begin{subequations}
\begin{equation}
k=\pi r-\frac{2\pi}{N}\sumir I_i,
\end{equation}
\begin{equation}
E=J\sumir\frac{-\sin^2\gam}{\ch 2z_i-\cos\gam}-h\(\frac{N}{2}-r\).
\end{equation}
\end{subequations}

Every solution of Bethe ansatz equations is uniquely determined
by Bethe quatum numbers $I_1<\dots<I_r$,
which provide us the classification of the excitations.
The ground state $|G\ket$ at magnetization $0\leq M_z\leq N/2$
is specified by the set of $r=N/2-M_z$ Bethe quantum numbers
$I_i=-N/4+M_z/2+i-1/2$.

\section{Dynamically dominant excitations}

In neutron scattering experiment at sufficiently low temperature,
the cross section is considered to be proportional to the dynamic spin structure factor
$S(q,\omega)$ at zero temperature,
which is defined by the space-time Fourier transform of the dynamical
correlation function:
\begin{eqnarray}
S_{\mu\bar{\mu}}(q,\omega)
&=&\frac{1}{N}\sum^N_{n,n'}e^{iq(n-n')}
  \int^{\infty}_{-\infty}dt\,e^{i\omega t}
  \bra S^{\mu}_n(t)S^{\bar{\mu}}_{n'}(0)\ket\nn\\
&=&\int^{\infty}_{-\infty}dt\,e^{i\omega t}\bra S^{\mu}_q(t)S^{\mu\dagger}_q(0)\ket\nn\\
&=&2\pi\sum_{\lam}|\bra G|S^{\mu}_q|\lam\ket|^2\delta(\omega-\omega_{\lam}),
\end{eqnarray}
where
$(\mu,\bar{\mu})=(z,z),(+,-),(-,+)$,
$\omega_{\lam}=E_{\lam}-E_G$.
$E_G$ and $E_{\lam}$ are the energy eigenvalues of
the ground state $|G\ket$ and
one of the excited states $|\lam\ket$,
respectively.
The spin fluctuation operator $S_q^{\mu}$ is defined by
\begin{equation}
S_q^{\mu}=\frac{1}{\sqrt{N}}\sumnN e^{iqn}S^{\mu}_n
\,,
\quad
\mu=z,+,-,
\end{equation}
for wave numbers $q=2\pi l/N$, $l=1,\dots,N$.

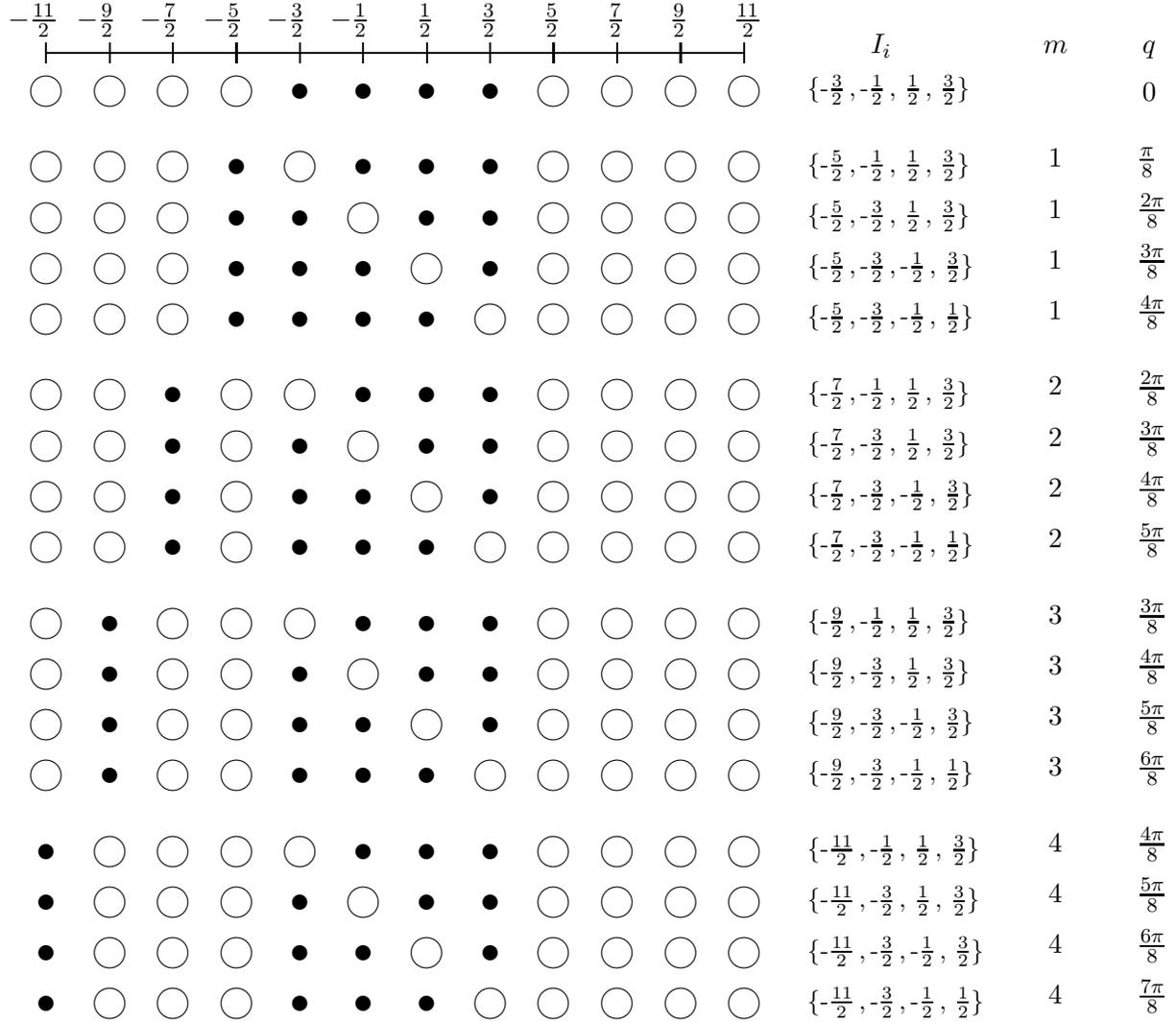
\begin{figure}
\label{configuration}
\begin{picture}(430,420)
\put(15,380){\line(1,0){275}}
\multiput(15,376)(25,0){12}{\line(0,1){8}}
\put(0,390){$-\frac{11}{2}$}
\put(27,390){$-\frac{9}{2}$}
\put(52,390){$-\frac{7}{2}$}
\put(77,390){$-\frac{5}{2}$}
\put(102,390){$-\frac{3}{2}$}
\put(127,390){$-\frac{1}{2}$}
\put(161,390){$\frac{1}{2}$}
\put(186,390){$\frac{3}{2}$}
\put(211,390){$\frac{5}{2}$}
\put(236,390){$\frac{7}{2}$}
\put(261,390){$\frac{9}{2}$}
\put(286,390){$\frac{11}{2}$}
\put(340,380){$I_i$}
\put(408,380){$m$}
\put(447,380){$q$}
\put(410,335){1}
\put(410,315){1}
\put(410,295){1}
\put(410,275){1}
\put(410,245){2}
\put(410,225){2}
\put(410,205){2}
\put(410,185){2}
\put(410,155){3}
\put(410,135){3}
\put(410,115){3}
\put(410,95){3}
\put(410,65){4}
\put(410,45){4}
\put(410,25){4}
\put(410,5){4}
\put(447,361){0}
\put(445,335){$\frac{\pi}{8}$}
\put(445,315){$\frac{2\pi}{8}$}
\put(445,295){$\frac{3\pi}{8}$}
\put(445,275){$\frac{4\pi}{8}$}
\put(445,245){$\frac{2\pi}{8}$}
\put(445,225){$\frac{3\pi}{8}$}
\put(445,205){$\frac{4\pi}{8}$}
\put(445,185){$\frac{5\pi}{8}$}
\put(445,155){$\frac{3\pi}{8}$}
\put(445,135){$\frac{4\pi}{8}$}
\put(445,115){$\frac{5\pi}{8}$}
\put(445,95){$\frac{6\pi}{8}$}
\put(445,65){$\frac{4\pi}{8}$}
\put(445,45){$\frac{5\pi}{8}$}
\put(445,25){$\frac{6\pi}{8}$}
\put(445,5){$\frac{7\pi}{8}$}
\put(15,365){\circle{12}}
\put(40,365){\circle{12}}
\put(65,365){\circle{12}}
\put(90,365){\circle{12}}
\put(115,365){\circle*{6}}
\put(140,365){\circle*{6}}
\put(165,365){\circle*{6}}
\put(190,365){\circle*{6}}
\put(215,365){\circle{12}}
\put(240,365){\circle{12}}
\put(265,365){\circle{12}}
\put(290,365){\circle{12}}
\put(315,363)
{\small{$\{\sm\frac{3}{2}\,,\,\sm\frac{1}{2}\,,\,\frac{1}{2}\,,\,\frac{3}{2}\}$}}
\put(15,335){\circle{12}}
\put(40,335){\circle{12}}
\put(65,335){\circle{12}}
\put(90,335){\circle*{6}}
\put(115,335){\circle{12}}
\put(140,335){\circle*{6}}
\put(165,335){\circle*{6}}
\put(190,335){\circle*{6}}
\put(215,335){\circle{12}}
\put(240,335){\circle{12}}
\put(265,335){\circle{12}}
\put(290,335){\circle{12}}
\put(315,333){\small{$\{\sm\frac{5}{2}\,,\,\sm\frac{1}{2}\,,\,\frac{1}{2}\,,\,\frac{3}{2}\}$}}
\put(15,315){\circle{12}}
\put(40,315){\circle{12}}
\put(65,315){\circle{12}}
\put(90,315){\circle*{6}}
\put(115,315){\circle*{6}}
\put(140,315){\circle{12}}
\put(165,315){\circle*{6}}
\put(190,315){\circle*{6}}
\put(215,315){\circle{12}}
\put(240,315){\circle{12}}
\put(265,315){\circle{12}}
\put(290,315){\circle{12}}
\put(315,313){\small{$\{\sm\frac{5}{2}\,,\,\sm\frac{3}{2}\,,\,\frac{1}{2}\,,\,\frac{3}{2}\}$}}
\put(15,295){\circle{12}}
\put(40,295){\circle{12}}
\put(65,295){\circle{12}}
\put(90,295){\circle*{6}}
\put(115,295){\circle*{6}}
\put(140,295){\circle*{6}}
\put(165,295){\circle{12}}
\put(190,295){\circle*{6}}
\put(215,295){\circle{12}}
\put(240,295){\circle{12}}
\put(265,295){\circle{12}}
\put(290,295){\circle{12}}
\put(315,293){\small{$\{\sm\frac{5}{2}\,,\,\sm\frac{3}{2}\,,\,\sm\frac{1}{2}\,,\,\frac{3}{2}\}$}}
\put(15,275){\circle{12}}
\put(40,275){\circle{12}}
\put(65,275){\circle{12}}
\put(90,275){\circle*{6}}
\put(115,275){\circle*{6}}
\put(140,275){\circle*{6}}
\put(165,275){\circle*{6}}
\put(190,275){\circle{12}}
\put(215,275){\circle{12}}
\put(240,275){\circle{12}}
\put(265,275){\circle{12}}
\put(290,275){\circle{12}}
\put(315,273){\small{$\{\sm\frac{5}{2}\,,\,\sm\frac{3}{2}\,,\,\sm\frac{1}{2}\,,\,\frac{1}{2}\}$}}
\put(15,245){\circle{12}}
\put(40,245){\circle{12}}
\put(65,245){\circle*{6}}
\put(90,245){\circle{12}}
\put(115,245){\circle{12}}
\put(140,245){\circle*{6}}
\put(165,245){\circle*{6}}
\put(190,245){\circle*{6}}
\put(215,245){\circle{12}}
\put(240,245){\circle{12}}
\put(265,245){\circle{12}}
\put(290,245){\circle{12}}
\put(315,243){\small{$\{\sm\frac{7}{2}\,,\,\sm\frac{1}{2}\,,\,\frac{1}{2}\,,\,\frac{3}{2}\}$}}
\put(15,225){\circle{12}}
\put(40,225){\circle{12}}
\put(65,225){\circle*{6}}
\put(90,225){\circle{12}}
\put(115,225){\circle*{6}}
\put(140,225){\circle{12}}
\put(165,225){\circle*{6}}
\put(190,225){\circle*{6}}
\put(215,225){\circle{12}}
\put(240,225){\circle{12}}
\put(265,225){\circle{12}}
\put(290,225){\circle{12}}
\put(315,223){\small{$\{\sm\frac{7}{2}\,,\,\sm\frac{3}{2}\,,\,\frac{1}{2}\,,\,\frac{3}{2}\}$}}
\put(15,205){\circle{12}}
\put(40,205){\circle{12}}
\put(65,205){\circle*{6}}
\put(90,205){\circle{12}}
\put(115,205){\circle*{6}}
\put(140,205){\circle*{6}}
\put(165,205){\circle{12}}
\put(190,205){\circle*{6}}
\put(215,205){\circle{12}}
\put(240,205){\circle{12}}
\put(265,205){\circle{12}}
\put(290,205){\circle{12}}
\put(315,203){\small{$\{\sm\frac{7}{2}\,,\,\sm\frac{3}{2}\,,\,\sm\frac{1}{2}\,,\,\frac{3}{2}\}$}}
\put(15,185){\circle{12}}
\put(40,185){\circle{12}}
\put(65,185){\circle*{6}}
\put(90,185){\circle{12}}
\put(115,185){\circle*{6}}
\put(140,185){\circle*{6}}
\put(165,185){\circle*{6}}
\put(190,185){\circle{12}}
\put(215,185){\circle{12}}
\put(240,185){\circle{12}}
\put(265,185){\circle{12}}
\put(290,185){\circle{12}}
\put(315,183){\small{$\{\sm\frac{7}{2}\,,\,\sm\frac{3}{2}\,,\,\sm\frac{1}{2}\,,\,\frac{1}{2}\}$}}
\put(15,155){\circle{12}}
\put(40,155){\circle*{6}}
\put(65,155){\circle{12}}
\put(90,155){\circle{12}}
\put(115,155){\circle{12}}
\put(140,155){\circle*{6}}
\put(165,155){\circle*{6}}
\put(190,155){\circle*{6}}
\put(215,155){\circle{12}}
\put(240,155){\circle{12}}
\put(265,155){\circle{12}}
\put(290,155){\circle{12}}
\put(315,153){\small{$\{\sm\frac{9}{2}\,,\,\sm\frac{1}{2}\,,\,\frac{1}{2}\,,\,\frac{3}{2}\}$}}
\put(15,135){\circle{12}}
\put(40,135){\circle*{6}}
\put(65,135){\circle{12}}
\put(90,135){\circle{12}}
\put(115,135){\circle*{6}}
\put(140,135){\circle{12}}
\put(165,135){\circle*{6}}
\put(190,135){\circle*{6}}
\put(215,135){\circle{12}}
\put(240,135){\circle{12}}
\put(265,135){\circle{12}}
\put(290,135){\circle{12}}
\put(315,133){\small{$\{\sm\frac{9}{2}\,,\,\sm\frac{3}{2}\,,\,\frac{1}{2}\,,\,\frac{3}{2}\}$}}
\put(15,115){\circle{12}}
\put(40,115){\circle*{6}}
\put(65,115){\circle{12}}
\put(90,115){\circle{12}}
\put(115,115){\circle*{6}}
\put(140,115){\circle*{6}}
\put(165,115){\circle{12}}
\put(190,115){\circle*{6}}
\put(215,115){\circle{12}}
\put(240,115){\circle{12}}
\put(265,115){\circle{12}}
\put(290,115){\circle{12}}
\put(315,113){\small{$\{\sm\frac{9}{2}\,,\,\sm\frac{3}{2}\,,\,\sm\frac{1}{2}\,,\,\frac{3}{2}\}$}}
\put(15,95){\circle{12}}
\put(40,95){\circle*{6}}
\put(65,95){\circle{12}}
\put(90,95){\circle{12}}
\put(115,95){\circle*{6}}
\put(140,95){\circle*{6}}
\put(165,95){\circle*{6}}
\put(190,95){\circle{12}}
\put(215,95){\circle{12}}
\put(240,95){\circle{12}}
\put(265,95){\circle{12}}
\put(290,95){\circle{12}}
\put(315,93){\small{$\{\sm\frac{9}{2}\,,\,\sm\frac{3}{2}\,,\,\sm\frac{1}{2}\,,\,\frac{1}{2}\}$}}
\put(15,65){\circle*{6}}
\put(40,65){\circle{12}}
\put(65,65){\circle{12}}
\put(90,65){\circle{12}}
\put(115,65){\circle{12}}
\put(140,65){\circle*{6}}
\put(165,65){\circle*{6}}
\put(190,65){\circle*{6}}
\put(215,65){\circle{12}}
\put(240,65){\circle{12}}
\put(265,65){\circle{12}}
\put(290,65){\circle{12}}
\put(315,63){\small{$\{\sm\frac{11}{2}\,,\,\sm\frac{1}{2}\,,\,\frac{1}{2}\,,\,\frac{3}{2}\}$}}
\put(15,45){\circle*{6}}
\put(40,45){\circle{12}}
\put(65,45){\circle{12}}
\put(90,45){\circle{12}}
\put(115,45){\circle*{6}}
\put(140,45){\circle{12}}
\put(165,45){\circle*{6}}
\put(190,45){\circle*{6}}
\put(215,45){\circle{12}}
\put(240,45){\circle{12}}
\put(265,45){\circle{12}}
\put(290,45){\circle{12}}
\put(315,43){\small{$\{\sm\frac{11}{2}\,,\,\sm\frac{3}{2}\,,\,\frac{1}{2}\,,\,\frac{3}{2}\}$}}
\put(15,25){\circle*{6}}
\put(40,25){\circle{12}}
\put(65,25){\circle{12}}
\put(90,25){\circle{12}}
\put(115,25){\circle*{6}}
\put(140,25){\circle*{6}}
\put(165,25){\circle{12}}
\put(190,25){\circle*{6}}
\put(215,25){\circle{12}}
\put(240,25){\circle{12}}
\put(265,25){\circle{12}}
\put(290,25){\circle{12}}
\put(315,23)
{\small{$\{\sm\frac{11}{2}\,,\,\sm\frac{3}{2}\,,\,\sm\frac{1}{2}\,,\,\frac{3}{2}\}$}}
\put(15,5){\circle*{6}}
\put(40,5){\circle{12}}
\put(65,5){\circle{12}}
\put(90,5){\circle{12}}
\put(115,5){\circle*{6}}
\put(140,5){\circle*{6}}
\put(165,5){\circle*{6}}
\put(190,5){\circle{12}}
\put(215,5){\circle{12}}
\put(240,5){\circle{12}}
\put(265,5){\circle{12}}
\put(290,5){\circle{12}}
\put(315,3)
{\small{$\{\sm\frac{11}{2}\,,\,\sm\frac{3}{2}\,,\,\sm\frac{1}{2}\,,\,\frac{1}{2}\}$}}
\end{picture}
\caption{
Configurations of Bethe quantum numbers of 
         psinon-antipsinon excitations for $N=16, M_z=4, r=4$.
$I_i$ are represented by the positions of small black circles.
Large circles represent $I_i$ vacancies.
The first row represents the ground state $|G\ket$.
}
\end{figure}

In the following, we will focus on the case for the parallel spin fluctuations
$(\mu,\bar{\mu})=(z,z)$,
$q=\pi/2$,
and for half the saturation magnetic field
$M_z=N/4$,
$r=N/2-M_z=N/4$.
In the case of XXX model ($\d=1$),
it was established that the spectral weight of $S_{zz}(q,\omega)$
is dominated by the set of collective excitations,
called "\textit{psinon}($\Psi$)-\textit{antipsinon}($\Psi^{\ast}$)" excitations,
which is denoted by $\Psi\Psi^{\ast}$\cite{line}.
Their Bethe quantum numbers are shown in figure \ref{configuration}.
The $I_i$ are given by the position of small black circles in each row.
The positions of large circles represent the $I_i$ vacancies.
The first row represents the ground state $|G\ket$,
whose Bethe quantum numbers are uniformly cofigured at the center.

When one of the Bethe quantum numbers of the ground state is moved to the
sea of $I_i$ vacancies, it becomes antipsinon.
Then it makes a hole in the $I_i$ configurations,
which represents the psinon.
In this way,
we have a set of excitations with two parameters,
which is called psinon-antipsinon excitations.
We label the $\p$ states with the integer parameter $m$.
The state $|m\ket$ has the Bethe quantum number $I_1=-r/2-m+1/2$.
%

The relative contribution of psinon-antipsinon excitations
is determined by
the ratio of the integrated intensity
\begin{equation}
S_{zz}(q)=\int^{\infty}_{-\infty}\frac{d\omega}{2\pi}S_{zz}(q,\omega)
=\sum_{\lam}|\bra G|S^z_q|\lam\ket|^2
=\bra G|S^z_qS^z_{-q}|G\ket
\end{equation}
and $\Psi\Psi^{\ast}$ contribution
\begin{equation}
S_{zz}^{\p}(q)=\sum_{\p}|\bra G|S^z_q|\p\ket|^2
=\sum_m|\bra G|S^z_q|m\ket|^2.
\end{equation}
Relative contributions 
for various values of $\d$ are shown in table \ref{cont}.
We see
they are more than
98\%
for general $\d$.
Although
the relative contribution actually decreases as the system size $N$ increases,
it was shown in Ref.[11] that
the relative contribution for the XXX model ($\d=1$)
is more than
$93\%$
from an extrapolation of the data for $N=12,16,20,24,28,32$.
%
%
%
%
Also we find that the relative contribution
is monotonously increasing
when we bring down the value of $\d$ from $1$ to $0$.
%
%
Hence we can conclude
that the $\p$ excitation is dynamically dominant
in the whole region $0<\d<1$.

\begin{table}
\caption
{Relative contributions of psinon-antipsinon excitations for $N=24$, $M_z=r=6$, $q=\pi/2$.}
\label{cont}
\begin{tabular}
{@{\hspace{\tabcolsep}\extracolsep{\fill}}cccc}
\hline
\hline
$\d$
&$S_{zz}^{\Psi\Psi^{\ast}}(\pi/2)$
&$S_{zz}(\pi/2)$
&Relative contribution(\%)\\
\hline
-0.4  & 0.228545 & 0.232882 & 98.1378 \\
-0.3  & 0.234930 & 0.237182 & 99.0504 \\
-0.1  & 0.245605 & 0.245782 & 99.9282 \\
-0.01 & 0.249581 & 0.249582 & 99.9994 \\
\hline
0.01  & 0.250415 & 0.250417 & 99.9994 \\
0.1   & 0.253984 & 0.254114 & 99.9489 \\
0.3   & 0.260983 & 0.261918 & 99.6431 \\
0.4   & 0.264076 & 0.265574 & 99.4360 \\
0.5   & 0.266954 & 0.269056 & 99.2189 \\
0.7   & 0.272230 & 0.275491 & 98.8161 \\
0.9   & 0.277048 & 0.281247 & 98.5068 \\
1     & 0.279318 & 0.283887 & 98.3907 \\
\hline
1.1   & 0.281503 & 0.286378 & 98.2977 \\
1.5   & 0.289411 & 0.295027 & 98.0965 \\
2.0   & 0.297553 & 0.303452 & 98.0559 \\
\hline\hline
\end{tabular}
\end{table}

\section{Transition rates and Dynamic spin structure factor}

We numerically evaluate the dynamic spin structure factor
from the contribution of $\p$ excitations:
\begin{equation}
S^{\p}_{zz}(q,\omega)=2\pi\sum_{m}|\bra G|S^z_q|m\ket|^2\delta(\omega-\omega_{m}).
\end{equation}
In the thermodynamic limit $N\rightarrow\infty$,
it can be represented by the product of the transition rates
$M_{zz}(q,\omega_m)=N|\bra G|S^z_q|m\ket|^2$
and the density of states
$D_{zz}(q,\omega_m)=\ds{\frac{2\pi}{N}\frac{1}{\omega_{m+1}-\omega_m}}$
\cite{quasi}.

From the determinant representations for the form factors and norms of
Bethe eigenstates\cite{kitanine},
we have the following formula for transition rates:
\begin{equation}
\label{formula}
\begin{split}
M_{zz}(q,\omega)
=&\frac{N^2}{4}\,\,
\frac{\ds{\prod^r_{j=1}\mid\sh(z^0_j-\ifgam)\mid^2}}
     {\ds{\prod^r_{j=1}\mid\sh(z_j-\ifgam)\mid^2}}\,\,
\left|\prod^r_{j<k}\frac{1}{\sinh^2(z^0_{j}-z^0_{k})+\sin^2\gam}\right|
\left|\prod^r_{\alpha<\beta}\frac{1}{\sinh^2(z_{\alpha}-z_{\beta})+\sin^2\gam}\right|\\
&\times
\frac{\mid \det(H-2P)\mid^2}{|\det{\it \Phi}(\{z^0_j\})|\,\,
|\det{\it \Phi}(\{z_j\})|}\,\,,
\end{split}
\end{equation}
where the matrix elements of $H$, $P$ and $\Phi$ are given by
\begin{equation}
\left\{
\begin{split}
&{\it \Phi}(\{z_j\})_{ab}=\left\{
\begin{array}{ll}
\ds{\frac{\sin 2\gam}{\sh^2(z_a-z_b)+\sin^2\gam}}&(a\neq b)\\[10pt]
\displaystyle{N\frac{\sin\gam}{\sh^2z_a+\sin^2\fgam}
-\sum_{\substack{k=1\\k\neq a}}^r\frac{\sin 2\gam}{\sh^2(z_a-z_k)+\sin^2\gam}}&(a=b)
\end{array}
\right.\\
&H(\{z^0_j\},\{z_k\})_{ab}
=\frac{1}{\sh(z^0_a-z_b)}
  \left(\prod^r_{j\neq a}\sh(z^0_j-z_b-i\gam)-
\left[\frac{\varph(z_b-\frac{i\gam}{2})}{\varph(z_b+\frac{i\gam}{2})}\right]^N
\prod^r_{j\neq a}\sh(z^0_j-z_b+i\gam)\right)\\
&P(\{z^0_j\},\{z_k\})_{ab}
=\frac{1}{\ds{\sh^2z^0_a+\sin^2\fgam}}\,\,\prod_{k=1}^r\sh(z_k-z_b-i\gam).
\end{split}
\right.
\end{equation}
$\{z^0_1,\dots,z^0_r\}$ and $\{z_1,\dots,z_r\}$
are the rapidities corresponding to
the ground state $|G\ket$
and one of the $\p$ states $|m\ket$,
respectively.
To obtain the desired transition rates,
we first solve the Bethe ansatz equations (\ref{bae})
for given Bethe quantum numbers $I_i$,
and then substitute the solution to the formula (\ref{formula}).

We rewrite the Bethe quantum numbers $I_i$ 
of the $\p$ state $|m\ket$ at $q=\pi/2$:
\begin{equation}
\begin{split}
I_1&=-r/2-m+1/2\\
I_i&=-r/2+i-3/2 \quad (2\leq i\leq r-m+1)\\
I_i&=-r/2+i-1/2 \quad (r-m+2\leq i\leq r)\\
   & \quad \quad \textrm{for}\quad 1\leq m\leq r
\end{split}
\end{equation}
When $m$ increases and approaches toward $r$,
the first Bethe quantum number $I_1$ gets away from 
the other $I_i$ configurations,
and the transition rate monotonously decreases.
At the same time it is getting more
difficult to find the solution of the first rapidity $z_1$.
%
%
%
%
%

\begin{figure}
\includegraphics{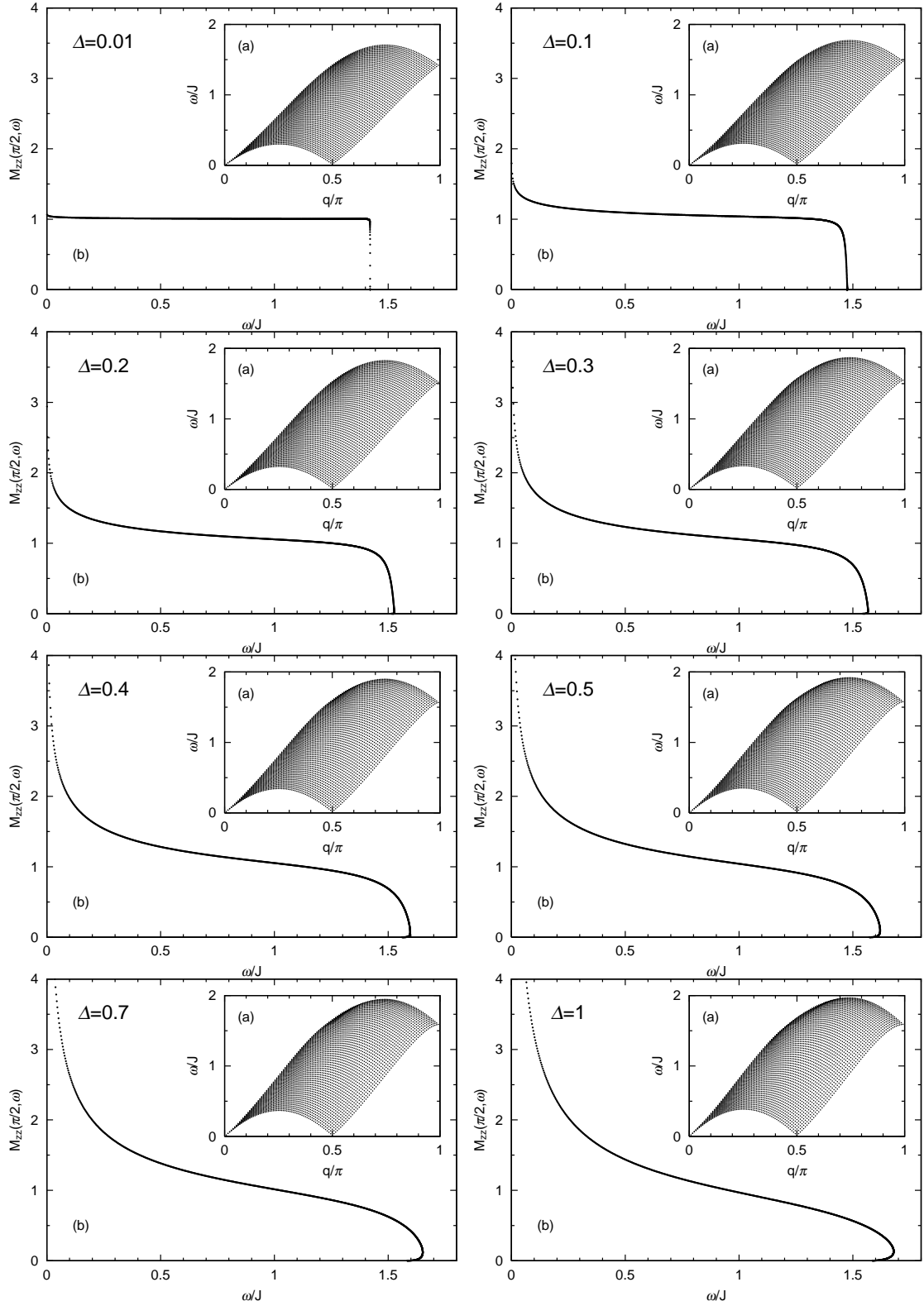}
\caption{
(a)Energy versus momentum of the psinon-antipsinon excitations
   at $M_z=N/4$ for $N$=256.
(b)Transition rates between the ground state and the psinon-antipsinon excitations
   at $q=\pi/2$, $M_z=N/4$ for $N$=4096.
}
\label{trbun}
\end{figure}

In figure \ref{trbun}(b),
we plot transition rates $M_{zz}(\pi/2,\omega)$ for $N=4096$.
Also shown in the inset (a) are
the excitation spectrum of $\p$ states for each $\d$.
Note that
in the XXX case ($\d=1$),
%
the transition rates $M_{zz}(\pi/2,\omega)$ can be observed as a smooth function
of $\omega$,
which has a singularity at the lower boundary,
and converges toward zero at the upper boundary.\cite{karxxx}

In the XXZ case
at the critical regime
$0<\d<1$,
it can be observed that
when we bring close to $\d=0$ starting from $\d=1$,
the singularity at the lower boundary has been weakened
and at last the function $M_{zz}(\pi/2,\omega)$
converges to the constant function with the value $1$.

The critical exponent of the  singularity at the lower boundary
can be compared with the prediction of conformal field theory.
The observed critical exponent $\eta_z$ in table \ref{crit} is a fit
$M_{zz}(\pi/2,\omega)=a+\omega^{\eta_z-2}$
from the data of the lowest and the second lowest excitations
$m=1,2$.
Conformal field theory predicts that it is equal to
the scaled energy gap
$2\theta_z=\Omega_z/(\pi v)$.\cite{crit}
$\Omega_z$ and the spin-wave velocity $v$ are defined by
the scaled lowest excitation energy
at $q=\pi/2$ and $q=2\pi/N$ in the thermodynamic limit:
\begin{equation}
\begin{split}
\Omega_z&=\lim_{N\rightarrow\infty}N\omega(m=1,q=\pi/2)\\
2\pi v&=\lim_{N\rightarrow\infty}N\omega(m=1,q=2\pi/N).
\end{split}
\end{equation}
They are shown for $0<\d\leq 1$ in table \ref{crit}.
We can see that the observed singularity and the prediction from
conformal field theory agree quite well.

\begin{table}
\caption{Critical Exponent of the singularity at the lower boundary}
\label{crit}
\begin{tabular}
{
ccccc}
\hline
\hline
$\Delta$    & 1 & 0.9 & 0.7 & 0.5  \\
\hline
$\Omega_z$  & 4.38286228 & 4.39050106 & 4.40585570 & 4.42060672 \\
$2\pi v$    & 5.72474288 & 5.64334684 & 5.45713287 & 5.23326627 \\
\hline
$2\theta_z$ & 1.53119969 & 1.55599192 & 1.61471447 & 1.68942549 \\
$\eta_z$    &   1.5313   &  1.5565    &  1.6163    &  1.6923    \\
\hline
\hline
\end{tabular}
\begin{tabular}
{
cccccc}
\hline
\hline
$\Delta$     & 0.4 & 0.3 & 0.2 & 0.1 & 0.01 \\
\hline
$\Omega_z$   & 4.42734708 & 4.43333735 & 4.43823777 & 4.44160837 & 4.44286904 \\
$2\pi v$     & 5.10437806 & 4.96243131 & 4.80595322 & 4.63334574 & 4.46278471 \\
\hline
$2\theta_z$  & 1.73472538 & 1.78676018 & 1.84697501 &1.91723589 & 1.99107478 \\
$\eta_z$     &  1.7384    &  1.7910    &  1.8516    &  1.9213    &  1.9918    \\
\hline
\hline
\end{tabular}
\end{table}

\begin{figure}
\includegraphics{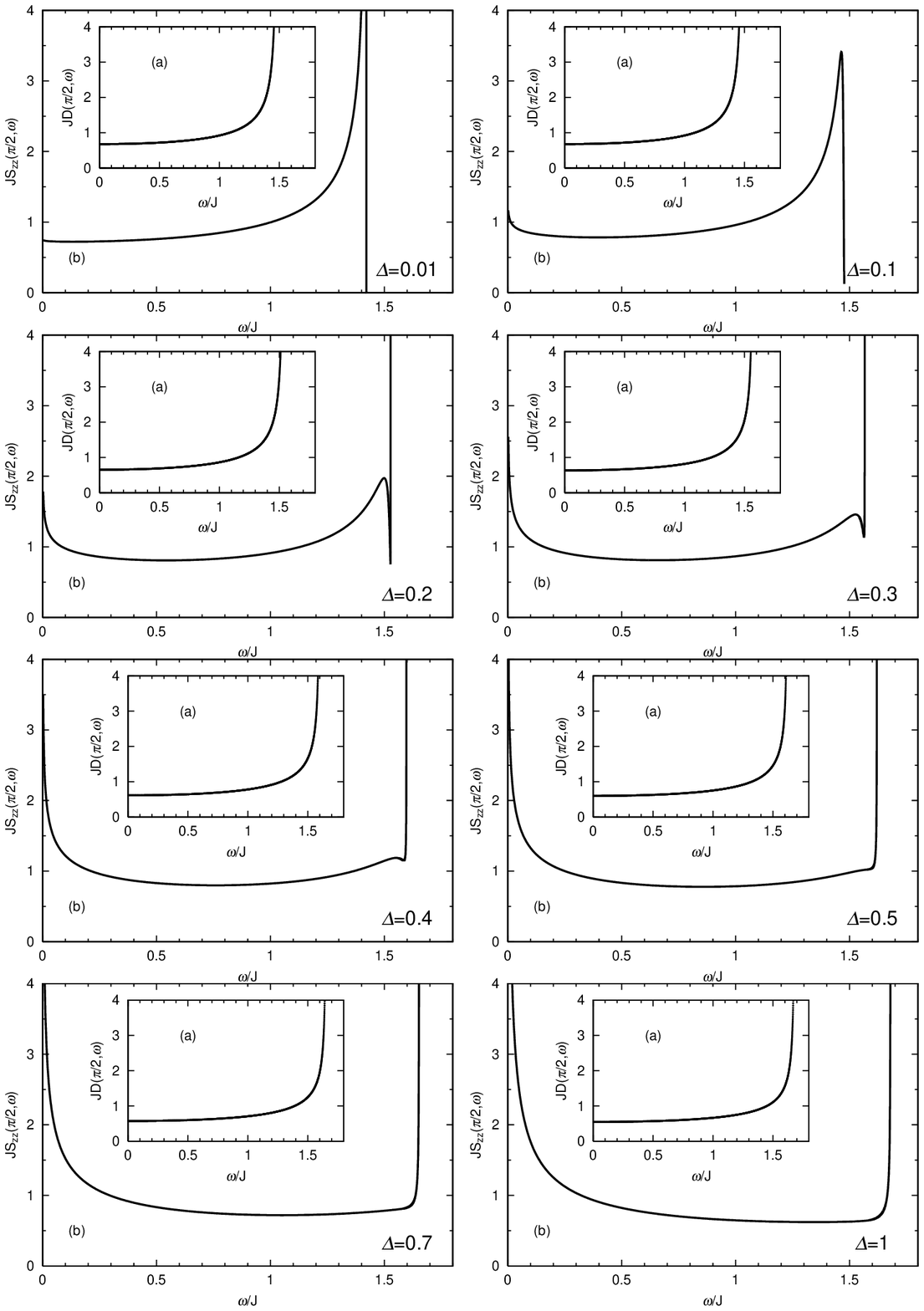}
\caption{
(a)Density of states for psinon-antipsinon excitations
         at $q=\pi/2$ , $M_z=N/4$ from data for $N$=4096.\\
(b)Spectral-weight distribution of psinon-antipsinon excitations in $S_{zz}(q,\omega)$
         at $q=\pi/2$ , $M_z=N/4$ from data for $N$=4096.
}
\label{sqd}
\end{figure}

Next we consider the dynamic spin structure factor $S_{zz}(\pi/2,\omega)$,
which is obtained by the product of transition rates $M_{zz}(\pi/2,\omega)$
and density of states $D_{zz}(\pi/2,\omega)$.
Density of states of $\p$ excitations are shown in figure \ref{sqd}(a).
It begins with flat function from the lower boundary,
and gradually increases toward the singularity at the upper boundary.

The spectral-weight distribution
$S_{zz}(\pi/2,\omega)$ is shown in figure \ref{sqd}(b).
In the XXX case $\d=1$, it has double singularities
 at the lower and upper boundary.\cite{karxxx}
Almost the same line shapes are observed for $0.7\leq\d<1$.
When we further decrease the value of $\d$,
a small peak near the upper boundary emerges for $\d\leq 0.5$,
and grows as
$\d$ is further decreasing.
In the limit $\d\rightarrow 0$,
this peak eventually becomes the singularity at the upper boundary,
where the line shape coincides with the density of states.

\section{Conclusion}

We have evaluated 
the longitudinal dynamic spin structure factor $S_{zz}(q,\omega)$
for the massless ($0<\d\leq1$) spin-1/2 XXZ chain
at $q=\pi/2$
with half the saturation magnetization $M_z=N/4$,
for the system size $N=4096$.
We have calculated the relative contribution of psinon-antipsinon excitations
and confirmed that they are also dominant in the critical regime.
We have also seen that
the observed critical exponents of the singularity at the lower boundary
agree with the prediction of conformal field theory.

We have observed here, for the first time,
that the $\d$-dependece of transition rates and
the spectral weight distribution
in the magnetic field.
It can be seen that the transition rates coverges toward the
constant function with the value $1$ in the limit $\d\rightarrow0$.
An interesting behavior of line shapes for $S_{zz}(\pi/2,\omega)$
is observed around $0.1<\d<0.5$,
where a small peak emerges near the upper boundary
and grows as $\d\rightarrow0$.
%
%
We hope this peculiar behavior of the peak will be measured
in inelastic neutron scattering experiments.

\section*{Acknowledgement}

We thank K. Sakai for useful discussions.
M. S. was supported by Grant-in-Aid for
Young Scientists No. 14740228.


\begin{thebibliography}{19}
%
%
\bibitem{Bethe} H.  Bethe, Z. Phys. {\bf 71}, 205 (1931).
%
%
\bibitem{Takabook} M. Takahashi,
                   \textit{Thermodynamics of One-Dimensional Solvable Models},
                   (Cambridge University Press, Cambridge, 1999).
%
%
\bibitem{cuno} P. R. Hammar, M. B. Stone, D. H. Reich, C. Broholm, P. J. Gibson, M. T. nad C. P. Landee,
               and M. Oshikawa, Phys. Rev. B {\bf 59}, 1008 (1999).
%
%
\bibitem{kcu} S. E. Nagler, D. A. Tennant, R. A. Cowley, T. G. Perring, and S. K. Satija, 
              Phys. Rev. B {\bf 44}, 12361 (1991). 
%
%
\bibitem{nmj} Th. Niemeijer, Physica {\bf 36}, 377 (1967).
%
%
\bibitem{stone} M. B. Stone, D. H. Reich, C. Broholm, K. Lefmann,
                C. Rischel, C. P. Landee and M. M. Turnbull,
                  Phys. Rev. Lett. {\bf 91}, 037205 (2003).
%
%
\bibitem{beck} G. M\"{u}ller, H. Thomas, H. Beck, and J. C. Bonner
                , Phys. Rev. B {\bf 24}, 1429 (1981).
%
%
\bibitem{bg} A. H. Bougourzi, M. Couture, and M. Kacir, Phys. Rev. B {\bf 54}, R12 669 (1996).
%
%
\bibitem{bgkar1} M. Karbach, G. M\"{u}ller, A. H. Bougourzi, A. Fledderjohann, and 
                 K. -H. M\"{u}tter, Phys. Rev. B {\bf 55}, 12 510 (1997).
%
%
\bibitem{bgkar2} A. H. Bougourzi, M. Karbach, G. M\"{u}ller, Phys. Rev. B {\bf 57}, 11 429 (1998).
%
%
\bibitem{jimbo} M. Jimbo and T. Miwa,
                \textit{Algebraic Analysis of Solvable Lattice Models},
                (American Mathematical Society, Providence, RI, 1994)
%
%
\bibitem{line} M. Karbach and G. M\"{u}ller, Phys. Rev. B {\bf 62}, 14 871 (2000).
%
%
\bibitem{quasi} M. Karbach, D. Biegel, and G. M\"{u}ller, Phys. Rev. B {\bf 66}, 054405 (2002).
%
%
\bibitem{kitanine} N. Kitanine, J. M. Maillet, and V. Terras, Nucl. Phys. B {\bf 554}, 647 (1999).
%
%
\bibitem{Korepinbook} V. E. Korepin, N. M. Bogoliubov and A. G. Izergin,
                   \textit{Quantum Inverse Scattering Method and Correlation Functions},
                   (Cambridge University Press, Cambridge, 1993).
\bibitem{karxxx} D. Biegel, M. Karbach, and G. M\"{u}ller
                  , Europhys. Lett {\bf 59}, 882 (2002).
%
%
\bibitem{karxxz} D. Biegel, M. Karbach, and G. M\"{u}ller
                  , J. Phys. A : Math. Gen. {\bf 36}, 5361 (2003).
%
%
\bibitem{karxx0} D. Biegel, M. Karbach, G. M\"{u}ller and K. Wiele,
                  , Phys. Rev. B {\bf 69}, 174404 (2004).
%
%
\bibitem{crit} A. Fledderjohann, C. Gerhardt, K. H. M\"{u}tter and A. Schmitt
                  , Phys. Rev. B {\bf 54}, 7168 (1996).
%
%
\end{thebibliography}
\end{document}